# Oersted field and spin current effects on magnetic domains in [Co/Pd]$_{15}$ nanowires

Pin Ho,[1*] Jinshuo Zhang,[1] David C. Bono,[1] Jingsheng Chen,[2] Caroline A. Ross[1]

[1]Department of Materials Science and Engineering, Massachusetts Institute of Technology, MA, USA
[2]Department of Materials Science and Engineering, National University of Singapore, Singapore, Singapore

**Abstract:**

An out-of-plane Oersted field produced from a current-carrying Au wire is used to induce local domain formation in wires made from [Co/Pd]$_{15}$ multilayers with perpendicular anisotropy. A 100 ns pulsed current of 56 – 110 mA injected into the Au wire created a reverse domain size of 120 – 290 nm in a Co/Pd nanowire on one side of the Au wire. A Biot-Savart model was used to estimate the position dependence of the Oersted field around the Au wire. The shape, size and location of the reversed region of Co/Pd was consistent with the magnitude of the Oersted field and the switching field distribution of the unpatterned film. A current density of $6.2 \times 10^{11}$ A m$^{-2}$ in the Co/Pd nanowire did not translate the domain walls due to low spin transfer efficiency, but the Joule heating promoted domain growth in a field below the coercive field.

*Corresponding author: P. Ho, E-mail: hopin@mit.edu

# I. INTRODUCTION

Domain wall (DW) motion in magnetic nanowires has been studied intensively due to its technological potential in achieving high density, high speed, low power computing logic devices and data storage devices such as racetrack memory.[1,2] Compared to materials with in-plane anisotropy, materials with perpendicular magnetic anisotropy (PMA) such as Co/Pt, Co/Ni, Co/Pd multilayers and CoFeB/MgO stacks have demonstrated or predicted performance improvements in terms of thermal stability, critical current, and switching speed.[3-5] The nucleation of reversed domains and the subsequent motion of DWs driven by a current is therefore of great interest.

In order to study current-induced DW motion in PMA nanowires, several methods have been adopted to introduce a reversed domain in a specific location, such as a thermo-magnetic writing process combining the use of a localised laser heating and reversed field,[6] or an injection pad with a smaller nucleation field than the nanowire.[7,8] A method that is compatible with device integration is the use of an Oersted field from a current-carrying wire orthogonal to the PMA nanowire to nucleate a local DW in the region of the PMA nanowire adjacent to the Au wire. There are several examples of this technique applied for instance to Co/Pt or Co/Ni with PMA of $(3.8 - 8.3) \times 10^6$ erg cm$^{-3}$.[9-13] In these experiments the Oersted field from the current pulses is chosen to be sufficient to produce a reversed domain, or in one case to control DW motion locally in a NiFe wire,[14] but the position and size of the reversed domain generated has not been examined as a function of current amplitude. Moreover, Oersted field-induced DW nucleation and DW motion are commonly detected by measuring the magneto-optic Kerr effect (MOKE) or anisotropic magnetoresistance, but these techniques lack the spatial resolution to determine domain shape and DW deformation[15] in the nanowire. Magnetic force microscopy (MFM) offers good spatial resolution (~50 nm) and is suitable for probing and understanding the domain shapes and deformation in nanowires with linewidths < ~200 nm.

In this article, we investigate the effects of the position-dependent Oersted field from a conductor wire on domain nucleation and location in PMA [Co/Pd]$_{15}$ nanostructures of different shapes and dimensions using MFM to image the domains. A model utilizing the Biot-Savart Law is proposed to estimate the Oersted field around a rectangular Au electrode and relate it to the positions of DWs produced by a current pulse. Finally, the effects of combined field and current on the expansion of the nucleated domains are studied.

## II. EXPERIMENTAL DETAILS

Straight and arc-shaped nanowires of linewidth 200 nm, circles of diameter ($d$) 2.5 μm and ovals with transverse ($d_y$) and longitudinal ($d_x$) diameter of 3 μm and 7 μm, respectively, were exposed in 160 nm thick polymethyl methacrylate resist spin coated on a naturally oxidized Si (100) substrate, using a Raith 150 electron beam lithography tool. [Co(0.6 nm)/Pd(1.2 nm)]$_{15}$ was subsequently deposited at room temperature on the patterned resist-coated Si/SiO$_2$ substrate by magnetron sputtering with Ar pressure of 10 mTorr and a base pressure better than $3 \times 10^{-7}$ Torr. The Co and Pd deposition rates with deposition power of 25 W were 0.62 and 1.86 nm min$^{-1}$, respectively. Top electrodes of Au (150 nm)/Ta (7 nm) were also patterned on the nanowire for the injection of the current pulse and electrical measurements. The as-grown [Co/Pd]$_{15}$ multilayer produced a strong MFM signal but the large anisotropy of the features required large current densities for Oersted field-induced DW formation. To reduce the perpendicular anisotropy, the patterned [Co/Pd]$_{15}$ film was post-annealed for 1 min at 150 °C, under a base pressure of $9 \times 10^{-7}$ Torr, to introduce mild interlayer diffusion. The patterned features and Au electrode pads were imaged by scanning electron microscopy (SEM) in the Raith 150. Magnetic properties of unpatterned [Co/Pd]$_{15}$ thin film were characterized by an ADE model 1660 vibrating sample magnetometer (VSM). Lattice spacing and crystal orientation were studied using X-ray diffraction (XRD) on a Panalytical multipurpose diffractometer with Cu K$_\alpha$ radiation. Feature dimensions were measured using a Veeco Nanoscope IV atomic force microscope (AFM). The magnetic domain structures in the wires were mapped out with MFM in the same instrument. Voltage pulses of magnitude 4 to 20 V and pulse widths of 100 ns to 1 μs were injected using a pulse generator (EH Research, Model 136A, 50 Ω).

## III. RESULTS AND DISCUSSION

### A. Unpatterned [Co/Pd]$_{15}$ film

The unpatterned [Co/Pd]$_{15}$ thin film shows a preferred (111) orientation (Fig. 1 inset) with a perpendicular magnetic anisotropy of 1.63 ($\pm$ 0.16) $\times$ 10$^6$ erg cm$^{-3}$, coercivity of 960 ($\pm$ 50) Oe and saturation magnetization 325 ($\pm$ 30) emu cm$^{-3}$ (Fig. 1). The magnetization reversal occurred over a field range of 800 ($\pm$ 50) Oe. The wide switching field distribution is attributed to local variability in the magnetic properties leading to DW nucleation and pinning. Variability can originate from non-uniform intermixing of the Co/Pd layers during post-annealing, from the film grain structure, and from nonuniform demagnetizing fields.

The field-induced reversal of patterned structures was probed by MFM. The 7 μm long oval shown in Fig. 2 was initially saturated 'down' with $H_z$ = -12 kOe. Upon applying a reversed field of $H_z$ = 400 Oe, reversed domains with 'up' configuration (yellow contrast) started to nucleate at the centre of the oval [Fig. 2 (a)]. With increasing reversed fields of $H_z$ = 500 to 700 Oe, domains propagated and additional domains nucleated [Fig. 2 (b)-(d)]. The fraction of the area that had reversed at -700 Oe was approximately 25% which is consistent with the VSM hysteresis loop of the unpatterned film.

**B. Effects of position-dependent Oersted field on [Co/Pd]$_{15}$ domains**

We next investigate the effects of a non-uniform Oersted field on patterned [Co/Pd]$_{15}$ patterned wires, circles and ellipses. The features are saturated with $H_z$ = 12 kOe to give an initial 'up' configuration. Fig. 3(a) shows a current-carrying Au wire patterned orthogonal to an arc-shaped nanowire with linewidth of 200 nm. The combined resistance of the circuit including the Au wire was 144 ± 5 Ω and the cross-sectional area of the Au wire was approximately 150 × 600 nm$^2$, therefore a voltage of magnitude 4 to 16 V generates a current of $I$ = 28 – 110 mA, corresponding to current density of $J$ = (3.1 – 12.3) × 10$^{11}$ A m$^{-2}$ in the Au wire. A short pulse width of 100 ns was used to minimize heat conduction from the Au wire into the [Co/Pd]$_{15}$ nanowire. When a current pulse of $I$ = 42 mA was injected through the Au wire from the top to bottom electrode, the MFM image [Fig. 3(b)] of the arc-shaped [Co/Pd]$_{15}$ nanowire did not show any change in the contrast, indicating that the Oersted field generated from the Au wire was insufficient to bring about local domain reversal at either side of the nanowire. At a current of $I$ = 56 mA, the [Co/Pd]$_{15}$ nanowire showed a slight change to a red contrast at the left region adjacent to Au wire [Fig. 3(c)], suggesting initial nucleation of a 'down' domain near the Au wire. With increased current densities to $I$ = 69 mA and 110 mA, the red contrast of the [Co/Pd]$_{15}$ nanowire to the left of the Au wire expanded [Fig. 3(d)-(e)]. This shows 'down' domain growth along the left part of the nanowire with increasing counter-clockwise Oersted field. The red contrast of the [Co/Pd]$_{15}$ nanowire to the immediate right of the Au wire [Fig. 3(c)-(e)], which was not present before current pulsing and at low currents [Fig. 3(b)], suggests that the reversed 'down' domain expanded under the Au wire. This may be a result of heat-assisted magnetization switching in the [Co/Pd]$_{15}$ beneath the Au wire. Joule heating of the Au wire will be discussed in later paragraphs.

The domain size for each current was obtained as the average of five measurements with respect to the left edge of the Au wire, after aligning the MFM images against a control

AFM image. Fig. 3(f) shows that the position of the DW with respect to the left edge of the Au wire increased from 120 to 290 nm when the current pulse increased from $I$ = 56 to 110 A.

The effect of a larger Oersted field in another [Co/Pd]$_{15}$ nanowire of linewidth 200 nm is shown in Fig. 4(a) after passing a current of 142 mA for 1 μs into the Au wire patterned orthogonal to the nanowire. The counter-clockwise Oersted field led to the formation of a reversed 'down' domain (red contrast) on the left of the Au wire [Fig. 4(b)]. Conversely, a clockwise Oersted field from a current pulse of opposite magnitude led to reversed domains on the right side of the Au wire [Fig. 4(c)]. As in Fig. 3(e), the transition between 'up' and 'down' domains was not straight, and small regions of reversed magnetization were present, indicating partial reversal over a region of the wire. The asymmetrical growth of the reversed domain on one side of the conductor is consistent with the opposite directions of the out-of-plane component of the field.

A model based on the Biot-Savart Law was used to estimate the Oersted field generated from the 150 × 600 nm$^2$ rectangular conductor [Fig. 5(a)]. The conductor is treated as 4 adjacent circular sub-wires of diameter 150 nm each carrying one quarter of the total current [Fig. 5(b)]. The total Oersted field at any position outside the conductor is a superposition of the fields contributed by all 4 circular sub-wires, each of which is given by $H = \frac{I_o}{2\pi R}$, where $I_o$ is the current in the sub-wire = $I/4$ and $R$ is the distance from the center of the sub-wire to the point of interest. For $I$ = 110 mA the model estimates an out-of-plane field of 1240 Oe at the edge of the Au wire, decreasing to 410 Oe at a distance of 290 nm from the wire edge [Fig. 5(c)].

The MFM and VSM results of Figs. 1 and 2 show that reversal in small regions occurs in fields as low as 400 Oe, and reversal is not complete until well above the coercive field of 960 Oe. The observations of reversed domains from the Oersted field are consistent with this range of switching fields. For example, when $I$ = 110 mA, the partially reversed region extends to ~300 nm corresponding to the onset of reversal at ~400 Oe. For $I$ = 56 mA, the Oersted field at the edge of the Au wire is calculated to be 620 Oe which is sufficient to cause nucleation of reversed domains as shown in Fig. 3(c).

The effects of an Oersted field generated from a current-carrying Au wire on the local DW formation in features of larger dimensions were also investigated, including circles and ovals (Fig. 6). The 150 × 600 nm² Au wire electrode patterned on top of the 2.5 µm diameter circle [Fig. 6(a)] had an overall resistance of 140 ± 5 Ω. A single 100 ns, 70 mA, 10 V pulse produced 'up' (yellow contrast) and 'down' magnetization (red contrast) on the left and right side of the Au wire, respectively [Fig. 6(b) and (c)]. Increasing the current density further to 85 mA increased the overall resistance of the Au wire to 199 ± 5 Ω suggesting damage to the conductor from Joule heating. A larger oval feature showed an increase in Au wire resistance for an even lower current of 65 mA [Fig. 6(e) and (f)].

The greater susceptibility of the Au wire to damage for increasing size of the underlying [Co/Pd]$_{15}$ features is attributed to the poorer heat dissipation of the section of Au wire in contact with the [Co/Pd]$_{15}$ thin film. The thermal diffusivity[16] of Co (26.8 ×10$^{-6}$ m² s$^{-1}$), Pd (24.7 ×10$^{-6}$ m² s$^{-1}$) or Co/Pd (10.0 ×10$^{-6}$ m² s$^{-1}$) is significantly lower than that of the Si substrate (8.6 ×10$^{-5}$ m² s$^{-1}$).[17, 18] The effect of SiO$_2$ on heat flow is neglected, given that it is a thin (~few nm) native oxide on the bulk Si substrate. The expected temperature change in the Au wire after a 100 ns pulse of 65 mA is calculated from the expression of You *et al.*[19] A temperature increase of 100-130 K is estimated for an Au wire in contact with bulk Co/Pd, and 39 K for Au on Si substrate, using the Au electrical conductivity of 1.0 × 10$^7$ Ω$^{-1}$ m$^{-1}$ (resistance: 54 Ω, Au wire length: 50 µm, cross-sectional area: 150 × 600 nm²), Co volumetric heat capacity of 3.7 × 10$^6$ J m$^{-3}$ K$^{-1}$,[20] Pd volumetric heat capacity of 2.9 × 10$^6$ J m$^{-3}$ K$^{-1}$,[20] Si volumetric heat capacity of 1.6 × 10$^6$ J m$^{-3}$ K$^{-1}$,[18] and assuming a Gaussian profile parameter of 0.5. The temperature rise calculated for Au on Co/Pd is likely an overestimate for the experimental situation of Au/[Co/Pd]$_{15}$/Si, but even so, we expect that when the size of the [Co/Pd]$_{15}$ feature increased, the poorer conduction of heat enhanced the temperature rise in the Au wire. In addition, reversed domains (bright features) are visible around both sides of the Au wire which is attributed to demagnetization in the heated regions. We assume the heating in the [Co/Pd]$_{15}$ lowers the switching field and enables propagation of the reversed domain under the Au as seen in the MFM images.

### C. Effects of field and spin current on [Co/Pd]$_{15}$ domains

We finally examine the effects of a current pulse applied to a [Co/Pd]$_{15}$ nanowire containing a reverse domain. Given the [Co/Pd]$_{15}$ nanowire resistance of 3000 $\pm$ 5 $\Omega$ and cross-sectional area of 27 $\times$ 200 nm$^2$, a pulse voltage of 10 V generates a current density of $J = 6.2 \times 10^{11}$ A m$^{-2}$ ($I = 3.3$ mA). A current pulse of this magnitude with duration 1 µs was injected into the [Co/Pd]$_{15}$ nanowire. This is estimated to lead to a temperature rise of 24 K in the [Co/Pd]$_{15}$ nanowire based on the [Co/Pd]$_{15}$ electrical conductivity of 1.2 $\times$ 10$^6$ $\Omega^{-1}$ m$^{-1}$ and the wire length between Au electrodes of 20 µm, the Si substrate thermal diffusivity and heat capacity stated earlier,[17,18] and assuming a Gaussian profile parameter of 0.5.[19] The current pulse alone had no detectable effect on the 'down' domain (red contrast) nucleated on the right of the Au wire [Fig. 4(c)]. However, the combination of an out-of-plane field $H_z$ = +550 Oe and $I$ = 3.3 mA promoted the growth of 'up' domains (yellow contrast) from the edge towards the centre, reducing the size of the reversed region [Fig. 4(d)]. Conversely, with $H_z$ = -550 Oe and $I$ = 3.3 mA, 'down' domains (red contrast) propagated from the edge towards the centre [Fig. 4(e)]. An applied field of +550 Oe alone produced reversed nucleation sites at the edge, but in smaller quantity and size compared to that observed under the combined influence of current and field.

These results suggest that current-driven domain wall motion was not significant in this regime of current density. Even though the walls are narrow (for [Co/Pd]$_{15}$ with a perpendicular $K_u$ of 1.63 ($\pm$ 0.16) $\times$ 10$^6$ erg cm$^{-3}$ the domain wall width $\delta = \pi\sqrt{A/K_u}$ where $A = 10^{-6}$ erg cm$^{-1}$ is estimated at $\delta = 25$ nm), and may therefore be expected to interact with high-frequency edge roughness, we believe the lack of current-driven DW motion is not an effect of the edge roughness of the wires, because the reversal fields in the large oval features and in the nanowires were similar based on the MFM measurements. Instead, it is likely that the spin transfer torque efficiency in [Co/Pd]$_{15}$ is low due to the presence of heavy metal scattering centres which lead to spin-independent scattering.[21] It is also most likely that the domain changes observed with combined field and current are promoted by Joule heating since there is no evidence of an unidirectionality of domain wall motion due to spin torque transfer. These results are consistent with Emori and Beach [22] and with prior work [11, 23] indicating only a small spin polarization of current in Co/Pt multilayers resulting from interface scattering.

## IV. SUMMARY

In conclusion, an out-of-plane Oersted field generated by a current-carrying Au wire orthogonal to a [Co/Pd]$_{15}$ nanowire was shown to induce a local reversed domain formation at one side of the Au wire. The size of the reversed region increased from 120 to 290 nm when the current in the Au wire increased from 56 to 110 mA. The irregular nature of the edge of the reversed domain is attributed to the gradually decreasing Oersted field and the switching field distribution of the film, and was consistent with the hysteresis measured from the unpatterned film. Passing a current of $6 \times 10^{11}$ A m$^{-2}$ through the [Co/Pd]$_{15}$ nanowire did not produce unidirectional domain wall motion indicating that the efficiency of spin transfer torque-driven domain wall motion in [Co/Pd]$_{15}$ was low. Instead, changes in domain configuration in the [Co/Pd]$_{15}$ were attributed to heating.

## V.   ACKNOWLEDGEMENTS

The authors gratefully acknowledge the support of the Agency of Science, Technology and Research (A*STAR) International Fellowship grant, and C-SPIN, a STARnet Center of the Semiconductor Research Corporation sponsored by DARPA and MARCO. Shared experimental facilities of CMSE, an NSF MRSEC under award DMR1419807, were used.

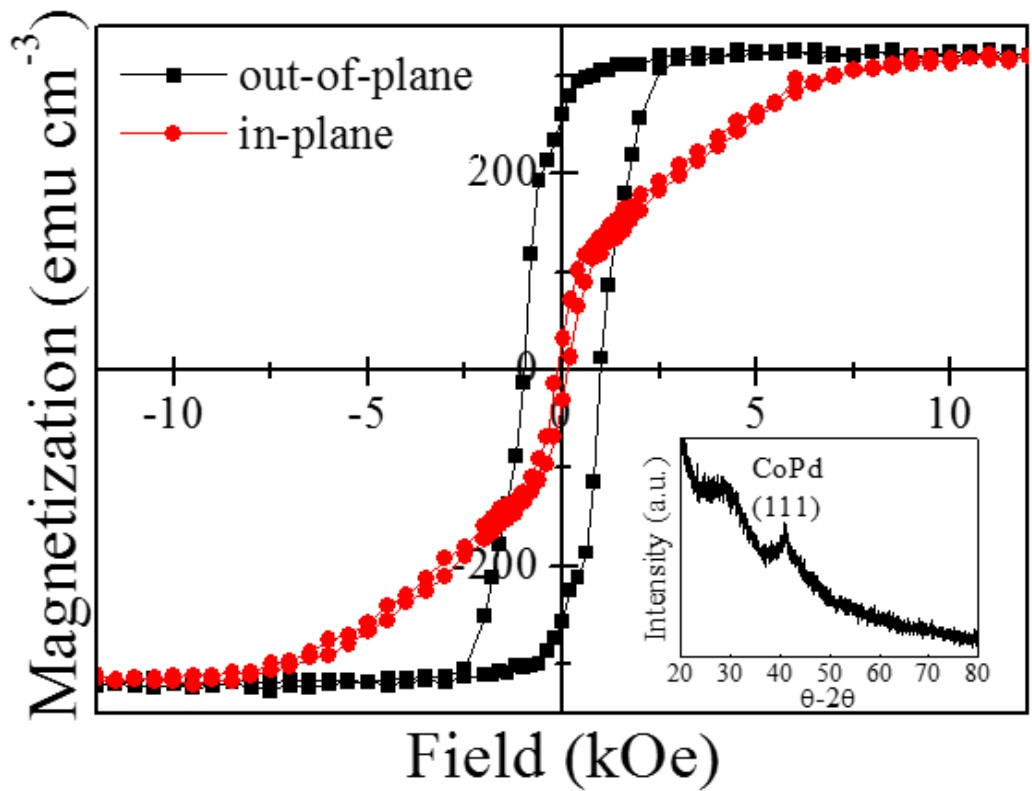

Figure 1. Hysteresis loop of the unpatterned Si/SiO$_2$/[Co(0.6 nm)/Pd(1.2 nm)]$_{15}$ thin film deposited at room temperature, followed by post-annealing treatment at 150 °C for one minute. Inset shows the XRD spectrum of the Si/SiO$_2$/[Co(0.6 nm)/Pd(1.2 nm)]$_{15}$ thin film.

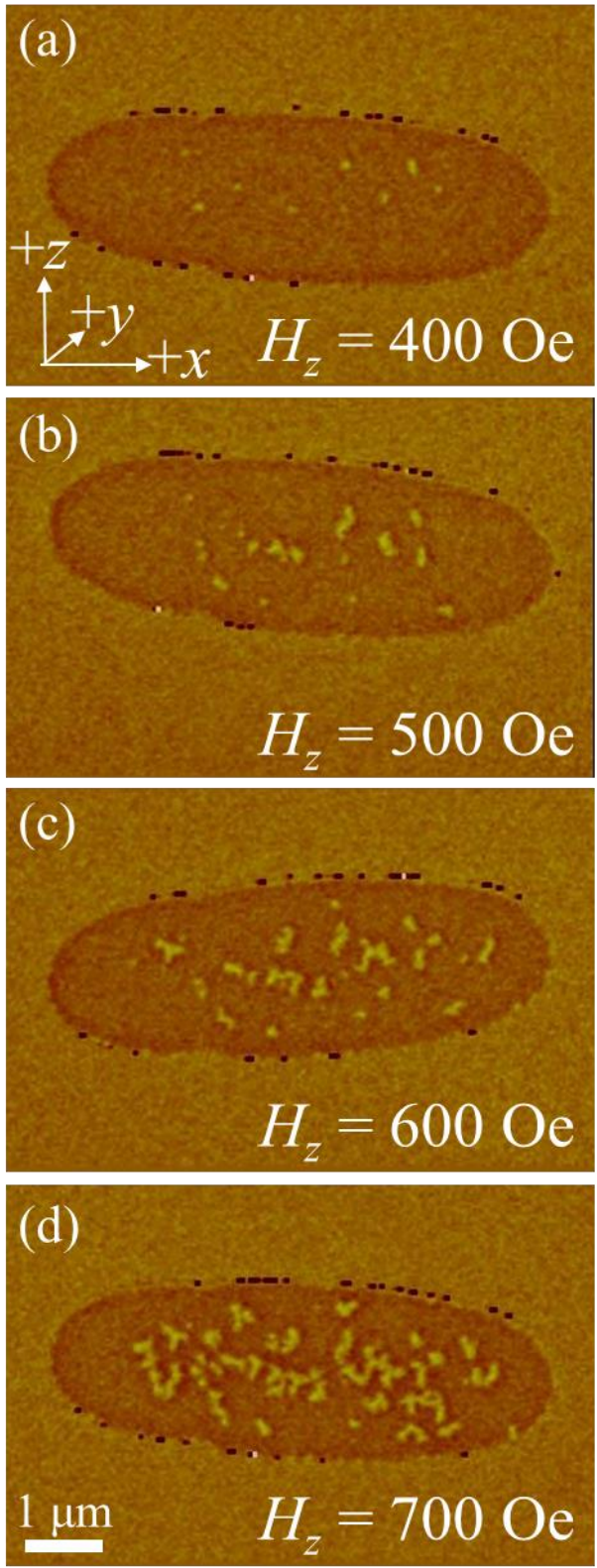

Figure 2. MFM images of a [Co/Pd]$_{15}$ oval feature at remanence. The sample was first saturated then a reversed field of $H_z$ = (a) 400, (b) 500, (c) 600 and (d) 700 Oe was applied.

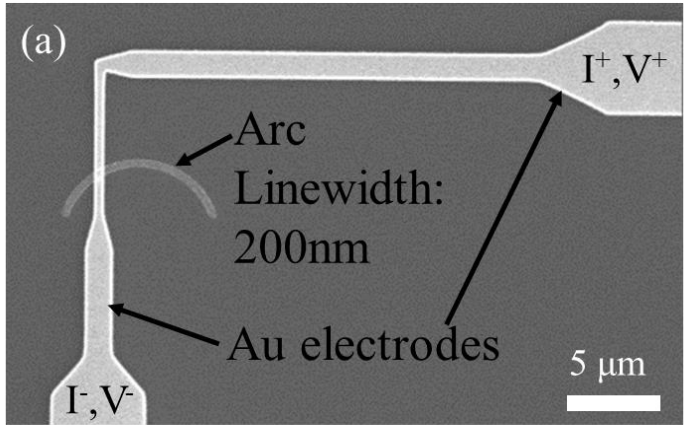
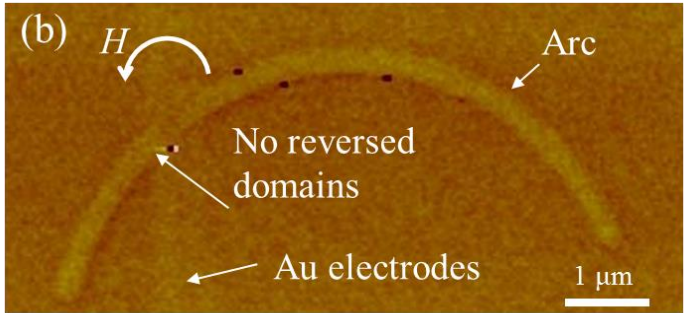
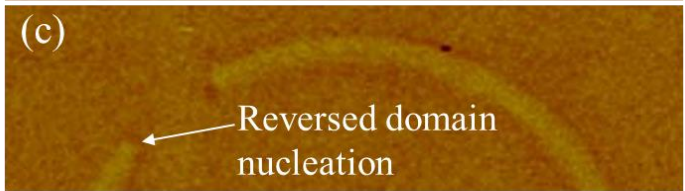
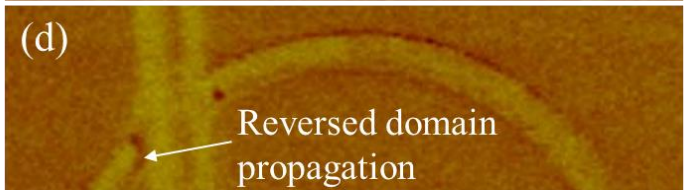
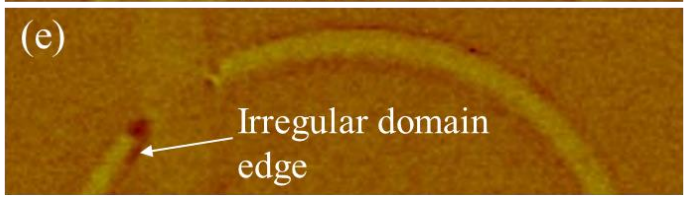
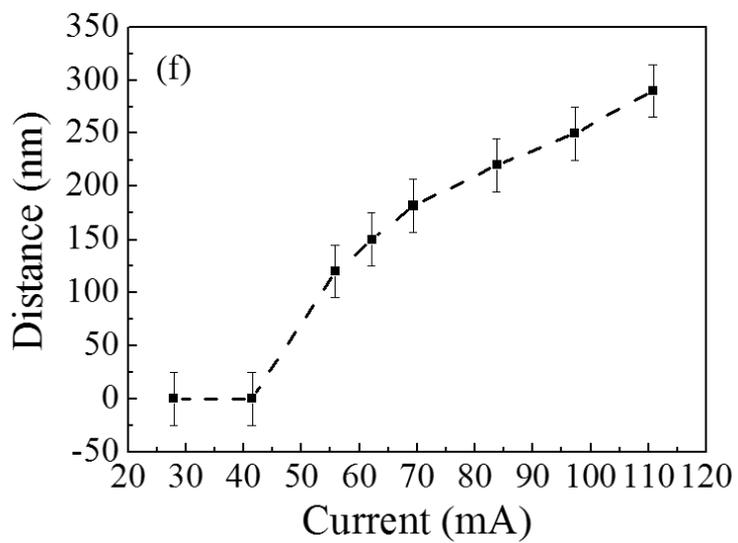

Figure 3. (a) SEM image showing a Au conductive wire placed on a [Co/Pd]$_{15}$ arc-shaped nanowire for the injection of a current pulse to generate an Oersted field. MFM images of the [Co/Pd]$_{15}$ nanowire illustrating DW formation at a current of (b) 42, (c) 56, (d) 69 and (e) 110 mA in the Au wire. (f) Average position of reversed DW from edge of Au wire as a function of current. The error bar in (f) is the MFM resolution estimated from the probe tip. The dashed lines serve as a guide for the eye.

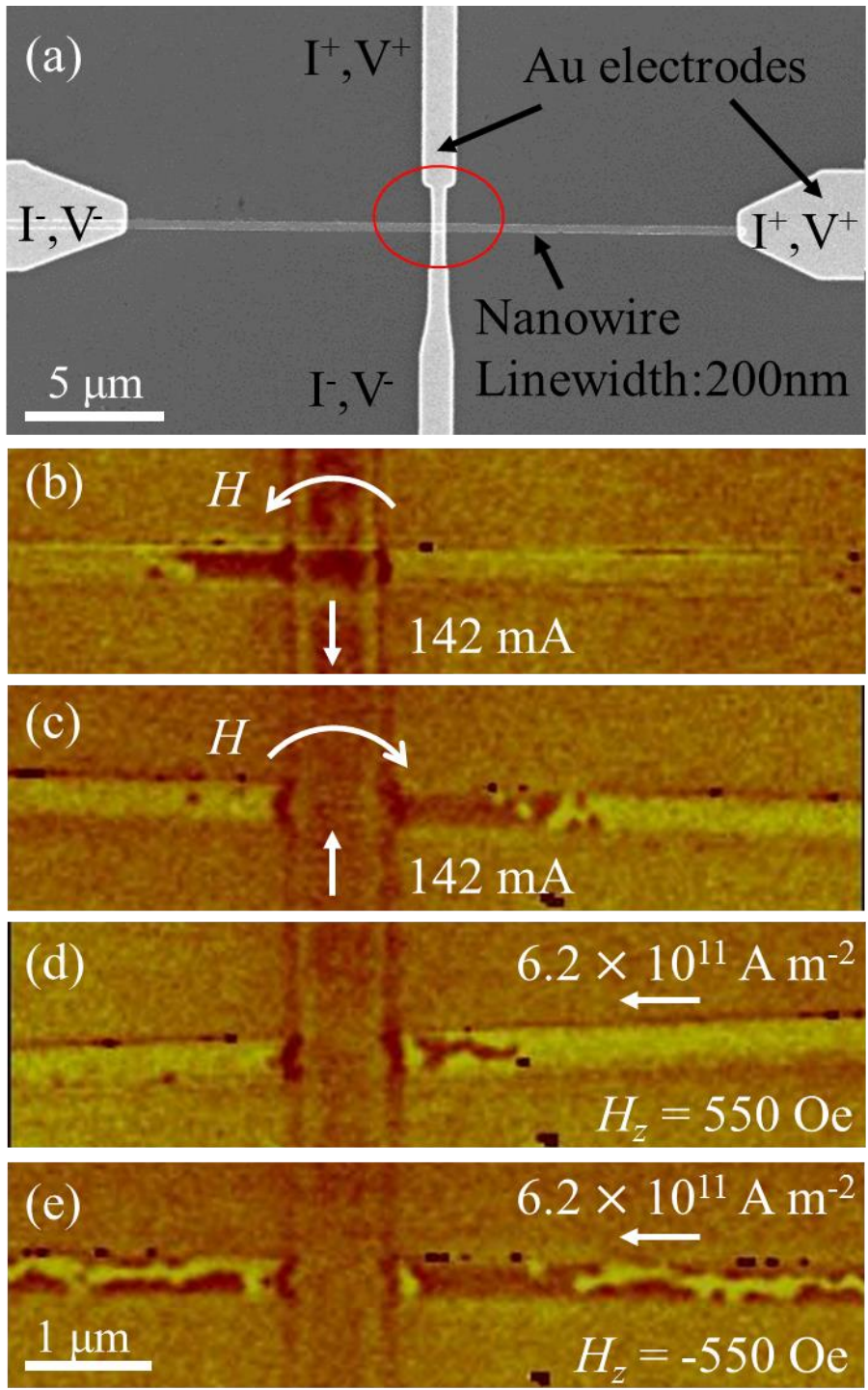

Figure 4. (a) SEM of [Co/Pd]$_{15}$ nanowire of linewidth 200 nm. MFM images of the [Co/Pd]$_{15}$ nanowire with the injection of a single 1 μs current pulse in the Au of (b) 142 mA generating a counter-clockwise Oersted field and (c) 142 mA generating a clockwise Oersted field. MFM images of the [Co/Pd]$_{15}$ nanowire demonstrating DW behaviour under the combined influence of a single 1 μs current density of $6.2 \times 10^{11}$ A m$^{-2}$ in the Co/Pd and (d) $H_z = 550$ Oe and (e) $H_z = -550$ Oe.

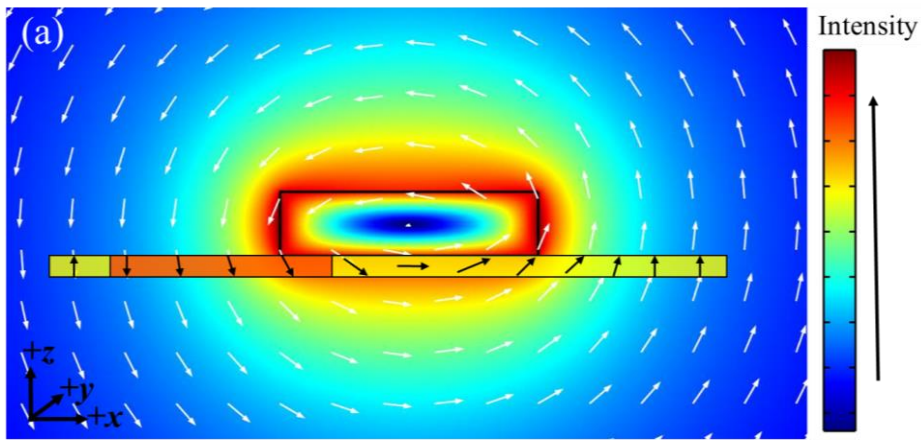

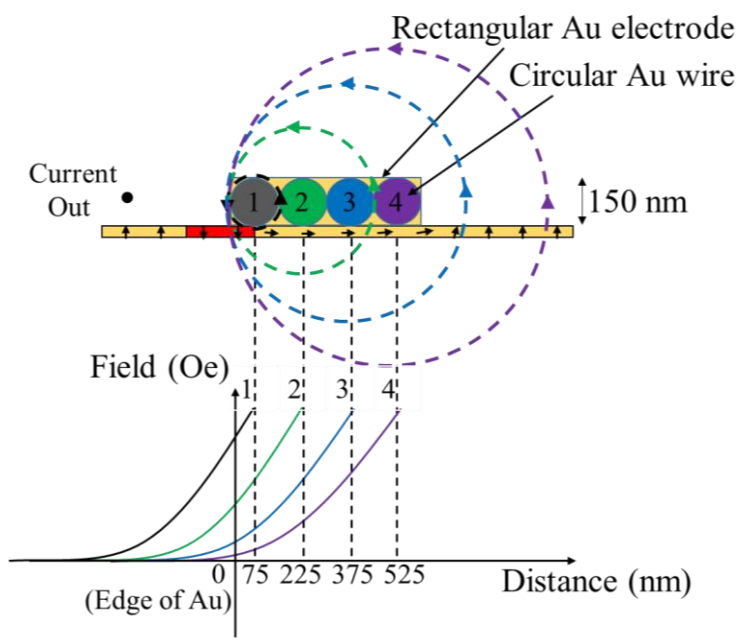

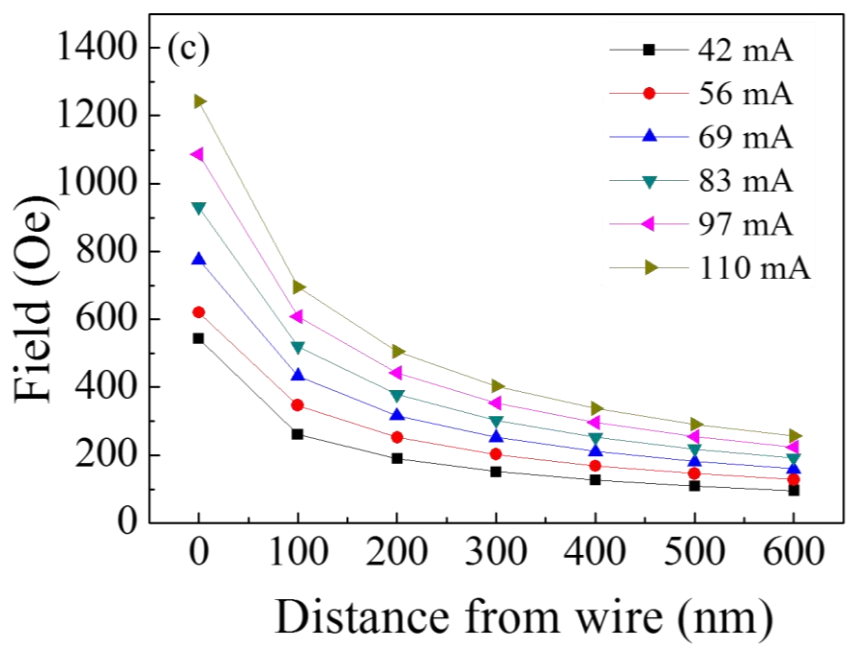

Figure 5. (a) Cross-sectional view of the rectangular 150 × 600 nm$^2$ Au conductor carrying an outward (along *y*-axis) current that generates a clockwise Oersted field influencing the magnetization of the perpendicular nanowire, shown schematically below the Au. (b) Diagram illustrating the model in which the rectangular 150 × 600 nm$^2$ Au electrode is treated as 4 adjacent circular sub-wires of diameter 150 nm each carrying an equal distribution of current. For a total current of 110 mA, the out-of-plane field at the edge of the Au wire is 1240 Oe (contribution from Wire 1: 740 Oe, Wire 2: 247 Oe, Wire 3: 148 Oe, Wire 4: 105 Oe). (c) Oersted field as a function of distance from the edge of the Au wire for different currents.

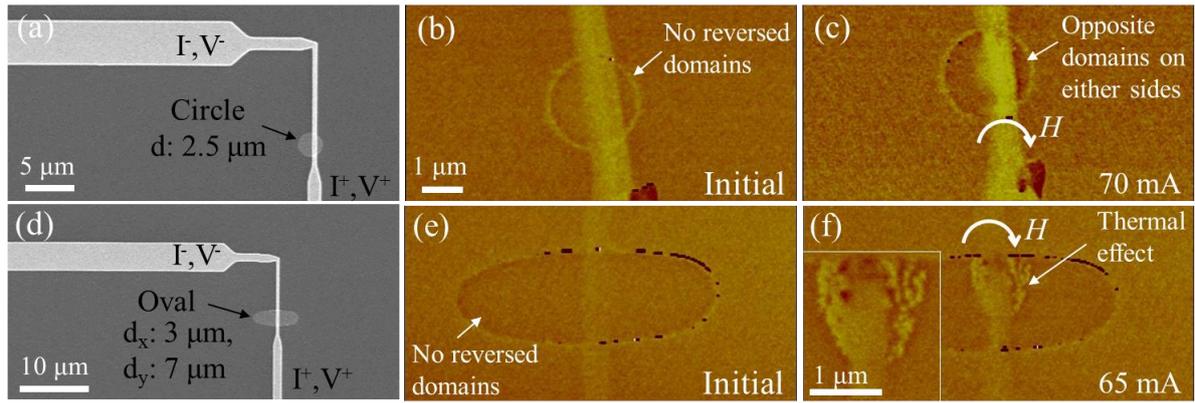

Figure 6. (a) SEM image showing the Au wire and electrode pads on the [Co/Pd]$_{15}$ circular feature for the injection of current pulse to generate an Oersted field. 10 × 5 μm MFM images of the [Co/Pd]$_{15}$ circle (b) before and (c) after injection of a single 100 ns pulse of current 70 mA. (d) SEM image showing the Au wire and electrode pads on the [Co/Pd]$_{15}$ oval feature. 10 × 5 μm MFM images of the [Co/Pd]$_{15}$ oval (e) before and (f) after injection of a single 100 ns pulse of current 65 mA. Inset is the magnification of the region showing domain formation due to Joule heating.